\documentstyle[12pt]{article}
\begin{document}        
\begin{titlepage}
\begin{flushright}
FERMILAB-PUB-97/263-T\\
NUHEP-TH-97-09\\
AMES-HET-97-08\\
July 1997
\end{flushright}
\vspace{0.1in}
\begin{center}
{\Large Single top quark production as a probe of R-parity-violating 
        SUSY \\
       at $pp$ and $p\bar p$ colliders}
\vspace{.2in}

  Robert J. Oakes$^a$, K. Whisnant$^b$, 
 Jin Min Yang$^{a,b,}${\footnote{ On leave from Physics 
Department, Henan Normal University, China}}
, Bing-Lin Young$^{b}$,  and X. Zhang$^{c, d}$

\vspace{.2in}
\it

$^a$     Department of Physics and Astronomy, Northwestern University,\\
         Evanston, Illinois 60208, USA \\
$^b$     Department of Physics and Astronomy, Iowa State University,\\
         Ames, Iowa 50011, USA\\    
$^c$     Institute of High Energy Physics, Academia Sinica, \\
         Beijing 100039, China\\
$^d$     Fermi National Accelarator Laboratory, P.O. Box 500, Batavia, IL 60510, USA
\rm
\end{center}
\vspace{3cm}

\begin{center} ABSTRACT\end{center}

 We investigate the ability of single top quark production via 
$q q' \rightarrow {\rm squark} \rightarrow t b$ and 
$q \bar q'\rightarrow {\rm slepton} \rightarrow t \bar b$
at the LHC and Tevatron to probe the strength of R-parity 
violating couplings in the minimal supersymmetric model.
We found that given the existing bounds on R-parity 
violating couplings,  single top quark production may be greatly
enhanced over that predicted by the standard model, and that
both colliders can either discover R-parity violating SUSY or set strong 
constraints on the relevant R-parity violating couplings.
We further found that the LHC is much more powerful 
than the Tevatron in probing the squark couplings,
but the two colliders have comparable sensitivity for the slepton 
couplings.
\vfill

\end{titlepage}
\eject
\baselineskip=0.30in
\begin{center} {\Large 1. Introduction }\end{center}

The HERA data showed excess events in deep-inelastic positron-proton
scattering at high-$Q^2$ and high $x$, which are in apparent conflict with the
Standard Model expectation [1]. The excess events have been interpreted 
as evidence of R-parity breaking supersymmetry[2]. Hence detailed 
examination of
effects of R-parity breaking supersymmetry in other processes is in order.
Some of the phenomenological implications of  R-parity violating couplings at 
$e^+e^-$ colliders have been investigated in Ref.[3]. 
Constraints on the R-parity violating couplings have also been obtained 
from perturbative unitarity [4,5],
$n-\bar n$ oscillation [5,6], 
$\nu_e$-Majorana mass [7], neutrino-less double $\beta$ decay [8], 
charged current universality [9], $e-\mu-\tau$ universality [9],
$\nu_{\mu}-e$ scattering [9], atomic parity violation [9], 
$\nu_{\mu}$ deep-inelastic scattering [9], double nucleon decay [10],
 $K$ decay [11,12],
$\tau$ decay [13], $D$ decay [13], $B$ decay [14-16] 
and $Z$ decay at LEPI [17,18].  Another important effect of 
R-parity violating couplings is that they may enhance the flavor 
changing top quark decays to the observable level of the upgraded 
Tevatron and LHC [19].

As is shown in Refs.[20-24], single top quark production  is very interesting 
to study at the Tevatron and the LHC since, in 
contrast to the QCD process of $t\bar t$ pair production,
it can be used to probe the electroweak theory. 
Single top production processes have been used to study the 
new physics effects involving the third-family quarks
in a model independent approach [25]
and in  specific models [26,27]. 
More recently, motivated by the evidence of R-parity breaking supersymmetry
[28,29] from the anomalous events at HERA [1], single top quark production 
$q\bar q'\rightarrow t\bar b$ at the Tevatron induced by baryon-number
violating 
(BV) couplings $\lambda''$ (via the exchange of a squark in the $t$-channel)
and by lepton-number violating (LV) couplings $\lambda'$ (via the exchange of a slepton 
in the $s$-channel) 
has been studied [30] in minimal supersymmetric model (MSSM).  
It was found [30] that the upgraded Tevatron 
can probe the relevant BV couplings 
efficiently, while the probe for the relevant LV couplings 
is very limited.

In addition to the process $q\bar q'\rightarrow t\bar b$ mentioned above, which
can be effectively studied at the Tevatron, the R-parity BV coupling
can lead to the reaction $q q'\rightarrow t b$
via an s-channel squark contribution which is suppressed at the Tevatron.
This process is suitable for study at the LHC and it probes a different 
set of BV couplings.
In this paper we make a detailed study of the s-channel BV effect. 
For completeness we study the effect at both the LHC and the upgraded
Tevatron.
We also study the s-channel slepton contribution to 
$q\bar q'\rightarrow t\bar b$ at the LHC 
which we compare to the result obtained for the upgraded Tevatron [30].

This paper is organized as follows.
In Sec.2 we present the Lagrangian for R-parity violating couplings
and squared matrix elements  for the processes 
$q q' \rightarrow {\rm squark} \rightarrow tb$  and
 $q\bar q' \rightarrow {\rm slepton}\rightarrow t\bar b$.
In Sec.3 we evaluate the signal for these processes and the SM background,
         and give the probing potential of the LHC in comparison 
         to the upgraded Tevatron.
\vspace{.5cm}

\begin{center} {\Large 2. $tb$ and  $t\bar b$ production 
                          in R-parity violating MSSM} \end{center}
\vspace{.5cm}

{\large 2.1 Lagrangian of  R-parity violating couplings}

The R-parity violating part of the superpotential of  the MSSM
is given by 
\begin{eqnarray}
{\cal W}_{\not \! R}&=&\lambda_{ijk}L_iL_jE_k^c
+\lambda_{ijk}^{\prime}L_iQ_jD_k^c
             +\lambda_{ijk}^{\prime\prime}U_i^cD_j^cD_k^c+\mu_iL_iH_2.
\end{eqnarray}
Here $L_i(Q_i)$ and $E_i(U_i,D_i)$ are the left-handed
lepton (quark) doublet and right-handed lepton (quark) singlet chiral 
superfields,
$i,j,k$ are generation indices, and $c$ denotes charge conjugation.
$H_{1,2}$ are the  chiral superfields
representing the two Higgs doublets. 
The $\lambda$ and $\lambda^{\prime}$ couplings 
violate lepton-number conservation, while the
$\lambda^{\prime\prime}$ couplings violate baryon-number conservation. 
The coefficient $\lambda_{ijk}$ is antisymmetric in the first two
indices and $\lambda^{\prime\prime}_{ijk}$ is antisymmetric in
the last two indices.
In terms of the four-component Dirac notation, the  Lagrangians for the
$\lambda^{\prime}$ and $\lambda^{\prime\prime}$ couplings that 
affect single top production at the Tevatron and the LHC are given by
\begin{eqnarray}
{\cal L}_{\lambda^{\prime}}&=&-\lambda^{\prime}_{ijk}
\left [\tilde \nu^i_L\bar d^k_R d^j_L+\tilde d^j_L\bar d^k_R\nu^i_L
       +(\tilde d^k_R)^*(\bar \nu^i_L)^c d^j_L\right.\nonumber\\
& &\hspace{1cm} \left. -\tilde e^i_L\bar d^k_R u^j_L
       -\tilde u^j_L\bar d^k_R e^i_L
       -(\tilde d^k_R)^*(\bar e^i_L)^c u^j_L\right ]+h.c.,\\
\label{eq3}
{\cal L}_{\lambda^{\prime\prime}}&=&-\lambda^{\prime\prime}_{ijk}
\left [\tilde d^k_R(\bar u^i_L)^c d^j_L+\tilde d^j_R(\bar d^k_L)^c u^i_L
       +\tilde u^i_R(\bar d^j_L)^c d^k_L\right ]+h.c..
\end{eqnarray}
The terms proportional to $\lambda$ are not relevant to our present 
discussion and will not be considered here.
Note that while it is theoretically possible to have both BV
and LV terms in the Lagrangian, the non-observation
of proton decay imposes very stringent conditions on their simultaneous
presence[31]. We, therefore, assume the existence of either  LV
couplings or BV couplings, and investigate  their separate 
effects in single top quark production.
\vspace{.5cm}

{\large 2.2 $q q' \rightarrow{\rm squark}\rightarrow tb$}

Production of $tb$ via an $s$-channel diagram 
$u^i d^j \rightarrow \tilde d^k_R \rightarrow tb$ can be induced by
the BV couplings $\lambda''$. The matrix element squared 
is given by 
\begin{equation}
\overline {\sum} \vert M^{ij}_{\lambda''}\vert^2=\frac{32}{3} 
\left \vert \sum_k \frac{\lambda''_{ijk}\lambda''_{33k}}
{\hat s-M^2_{\tilde d^k}+iM_{\tilde d^k}\Gamma_{\tilde d^k_R}} \right\vert^2
(p_1\cdot p_2)\left [ p_3\cdot p_4-M_t (s_t\cdot p_4)\right ],
\end{equation}
where $p_1$ and $p_2$ denote the momenta of the incoming quarks $u^i$ 
and $d^j$, $p_3$ and $p_4$ of the outgoing $t$ and
$b$ quarks.  The center-of-mass energy of the parton is given by $\hat{s}$
and $s_t$ denotes the spin of top quark which is given by
\begin{equation}
s_t=\frac{h}{M_t}\left (\vert \vec p_3 \vert, E_t \hat p_3\right ),
\end{equation}
where $h=\pm 1$ denotes the two helicity states, and $\hat p_3$ is the unit 
three-vector in the momentum direction of top quark. 

Neglecting the contribution of third-family sea quark in the
initial states, we obtain
\begin{eqnarray}
\overline {\sum }\vert M_{\lambda''}(ud\rightarrow tb) \vert^2&=&\frac{32}{3} 
 \frac{(\lambda''_{112}\lambda''_{332})^2}
{(\hat s-M^2_{\tilde s})^2+(M_{\tilde s}\Gamma_{\tilde s_R})^2 } 
(p_1\cdot p_2)\left [ p_3\cdot p_4-M_t (s_t\cdot p_4)\right ],\\
\overline {\sum} \vert M_{\lambda''}(us\rightarrow tb) \vert^2&=&\frac{32}{3} 
\frac{(\lambda''_{112}\lambda''_{331})^2}
{(\hat s-M^2_{\tilde d})^2+(M_{\tilde d}\Gamma_{\tilde d_R})^2} 
(p_1\cdot p_2)\left [ p_3\cdot p_4-M_t (s_t\cdot p_4)\right ],\\
\overline {\sum} \vert M_{\lambda''}(cd\rightarrow tb) \vert^2&=&\frac{32}{3} 
\frac{(\lambda''_{212}\lambda''_{332})^2}
{(\hat s-M^2_{\tilde s})^2+(M_{\tilde s}\Gamma_{\tilde s_R})^2}
(p_1\cdot p_2)\left [ p_3\cdot p_4-M_t (s_t\cdot p_4)\right ],\\
\overline {\sum} \vert M_{\lambda''}(cs\rightarrow tb) \vert^2&=&\frac{32}{3} 
\frac{(\lambda''_{212}\lambda''_{331})^2}
{(\hat s-M^2_{\tilde d})^2+(M_{\tilde d}\Gamma_{\tilde d_R})^2} 
(p_1\cdot p_2)\left [ p_3\cdot p_4-M_t (s_t\cdot p_4)\right ].
\end{eqnarray}

In the R-parity conserving MSSM, the down-type squark $\tilde d^k_R$ 
can decay into charginos and neutralinos via the processes
$\tilde d^k_R\rightarrow u^k+\overline{\tilde \chi^+_j}$ ($j=1,2$) and
$\tilde d^k_R\rightarrow d^k+\tilde \chi^{0}_j$ ($j=1,2,3,4$),
where $\tilde \chi^+_j$ and $\tilde \chi^0_j$
represent a chargino and neutralino, respectively [32].
Of course, it can also decay into gluino plus quark if kinematically allowed.
In the R-parity violating MSSM,  the down-type squark $\tilde d^k_R$
can also decay into quark pairs $\tilde d^k\rightarrow \bar d^j +\bar u^i$
via the $\lambda^{\prime\prime}$ couplings.
Since some of the relevant $\lambda^{\prime\prime}$ couplings may be
quite large, the width of a heavy down-type squark $\tilde d^k_R$ can be 
large even if we do not consider the decay involving the gluino.
We found that within the allowed parameter space ($\lambda''$, chargino 
and neutralino sector) its width $\Gamma_{\tilde d_R}$
can be as large as $M_{\tilde d_R}/3$.    
\vspace{.5cm}

{\large 2.3  $q\bar q' \rightarrow{\rm slepton}\rightarrow t\bar b$}

Production of $t\bar b$ via an $s$-channel slepton 
$u^i \bar d^j \rightarrow \tilde e^k_L \rightarrow t\bar b$  can be 
induced by the LV couplings $\lambda'$. The matrix element squared 
is given by 
\begin{equation}
\overline {\sum} \vert M^{ij}_{\lambda'}\vert^2=\frac{1}{2} 
\frac{ \left (\lambda'_{1ij}\lambda'_{133}+\lambda'_{2ij}\lambda'_{233}
+\lambda'_{3ij}\lambda'_{333}\right )^2}
{(\hat s-M^2_{\tilde e_L})^2+(M_{\tilde e_L}\Gamma_{\tilde e_L})^2}
(p_1\cdot p_2)\left [ p_3\cdot p_4-M_t (s_t\cdot p_4)\right ],
\end{equation}
where we assumed mass degeneracy for sleptons of different flavors.

In the R-parity conserving MSSM, the charged sleptons $\tilde e_L$ will
decay into charginos and neutralinos via the processes
$\tilde e_L\rightarrow \nu_{e}+\overline{\tilde \chi^+_j}$ ($j=1,2$) and
$\tilde e_L\rightarrow e+\tilde \chi^{0}_j$ ($j=1,2,3,4$) [32].
However, in the R-parity violating MSSM, the slepton can also decay 
into quark pairs via the $\lambda^{\prime}$ couplings
$\tilde e^i_L\rightarrow \bar u^j_L+d^k_R$. 
Since the allowed $\lambda^{\prime}$ couplings
are quite small, the dominant decays are the chargino and neutralino modes.
The partial widths are given by
\begin{eqnarray}
\Gamma(\tilde e_L\rightarrow \nu_{e}+\overline{\tilde \chi}^+_j)
&=&\frac{g^2}{16\pi M_{\tilde e}^3}\left \vert U_{j1}\right \vert^2
\left (M_{\tilde e}^2-M_{\tilde \chi^+_j}^2\right )^2,\\
\Gamma(\tilde e_L\rightarrow e+\tilde \chi^{0}_j)
&=&\frac{g^2}{8\pi M_{\tilde e}^3}
\left\vert s_W N^{\prime}_{j1}
+\frac{1}{c_W}(\frac{1}{2}-s_W^2)N^{\prime}_{j2}\right\vert^2
\left (M_{\tilde e}^2-M_{\tilde \chi^0_j}^2\right )^2,
\end{eqnarray}
where $s_W\equiv \sin\theta_W, c_W\equiv \cos\theta_W$
and the masses of the lepton and down-type quarks are neglected.
The masses of charginos and neutralinos, and the matrix elements
$U_{ij}$ and $N^{\prime}_{ij}$ which respectively diagonalize the mass
 matrix of chargino and neutralino, depend on the SUSY parameters $M_2$, $M_1$,
$\mu$, and $\tan\beta$ [29].
Here, $M_2$ and $M_1$ are the masses of gauginos corresponding to
$SU(2)$ and $U(1)$, respectively,  $\mu$ is the coefficient of
the $H_1H_2$ mixing term in the superpotential,
and $\tan\beta=v_2/v_1$ is the ratio of the
vacuum expectation values of the two Higgs doublets.
\vspace{.5cm}

\begin{center} {\Large 3. Numerical calculation and results } \end{center}
\vspace{.5cm}

Due to the large QCD backgrounds at hadron colliders, it is very difficult,
if not impossible, to search for the signal from the hadronic decays of
the top quark.  We therefore look for events with 
$t\rightarrow W^+b \rightarrow l^+\nu b$ ($l=e, \mu$). 
(We take into account of the fact that the top quark is polarized in
hadronic production.)
 Thus, the signature of this process is an energetic
charged lepton, missing $E_{T}$, and  double $b$-quark jets.
We assumed silicon vertex tagging of the $b$-quark jet 
with $50 \%$ effeciency and the probability of 0.4\% for a light quark 
jet to be mis-identified as a $b$-jet [33]. 

Although the present events have the unique signal of same
sign b-quarks, since the tagging can not distinguish a b-quark 
jet from $\bar b$-quark jet, there are many potential SM backgrounds [33]:
\begin{itemize}
\begin{description}
\item[{\rm(1)}] the Drell-Yan like process 
                $q\bar q' \rightarrow W^* \rightarrow t\bar b$;
\item[{\rm(2)}] the quark-gluon process  
                $qg\rightarrow q't\bar b$ with a W-boson
                as an intermediate state in either the t-channel 
                or the s-channel of a subdiagram;
\item[{\rm(3)}] processes involving a b-quark in the initial state,
                $bq (\bar q) \rightarrow tq'(\bar q')$ and $gb\rightarrow tW$;
\item[{\rm(4)}] $Wb\bar b$;
\item[{\rm(5)}] $Wjj$;
\item[{\rm(6)}] $t\bar t\rightarrow W^-W^+b\bar b$.
\end{description}
\end{itemize}
 Background process (2) contains an
extra quark jet and can only mimic our signal 
if the quark misses detection by going into the beam 
pipe. This can only happen when the light quark jet has 
 the pseudorapidity greater than about 3 or the transverse momentum less than 
about 10 GeV.
In our calculation of the $W$-gluon fusion process as a background, we
impose $\eta(q') >3$ and $p_T(q')< 10$ GeV for the  light-quark jet.
The $bq (\bar q) \rightarrow tq'(\bar q')$ background is greatly 
reduced by requiring double b-tagging. The process $gb\rightarrow tW$
can only imitate our signal if the $W$ decays into two jets, where 
one jet is missed by the detector and the other is mis-identified as a $b$ 
quark, which should be negligible. 
Since we required two $b$-jets to be present in the final state
and assumed the probability for a light quark 
jet to be mis-identified as a $b$-jet is 0.4\%,
the potentially large background process (5) from $Wjj$ is reduced 
to an insignificant level.
Also we required the reconstructed top quark mass $M(bW)$ to lie 
within the mass range 
\begin{equation}
\left \vert M(bW)-m_t\right \vert <30~{\rm GeV},
\end{equation}
which can also reduce the backgrounds $Wb\bar b$ and $Wjj$ efficiently.
Background process (6) can mimic our signal if both $W$'s decay leptonically
and one charged lepton is not detected, which we assumed to occur if
$\eta(l)>3$ and $p_T(l)<10$ GeV.

To make a realistic estimate  we also need to 
consider the detector acceptance.
 To simulate the detector acceptance,  we made a series of
cuts on the transverse momentum ($p_{T}$), the pseudo-rapidity
($\eta$), and the separation in the azimuthal angle-pseudo rapidity plane 
(~$\Delta R= \sqrt{(\Delta \phi)^2 +
(\Delta \eta)^2}~ )$ between a jet and a lepton or between two jets.
For the upgraded Tevatron, the cuts are chosen to be
\begin{eqnarray}
p_T^l, ~p_T^b,~
p_T^{\rm miss}&\ge& 20 \rm{~GeV} ~,\\
\eta_{b},~\eta_{l} &\le& 2.5 ~,\\
\Delta R_{jj},~\Delta R_{jl} &\ge& 0.5 ~.
\end{eqnarray}
For the LHC,  the cuts are chosen to be
\begin{eqnarray}
p_T^l&\ge& 20 \rm{~GeV} ~,\\
p_T^b&\ge& 35 \rm{~GeV} ~,\\
p_T^{\rm miss}&\ge& 30 \rm{~GeV} ~,\\
\eta_{b},~\eta_{l} &\le& 3 ~,\\
\Delta R_{jj},~\Delta R_{jl} &\ge& 0.4~.
\end{eqnarray}

To make the analyses more realistic, we simulate the detector effects
by  assuming a Gaussian smearing from the energy of the final state
particles, given by:
\begin{eqnarray}
\Delta E / E & = & 30 \% / \sqrt{E} \oplus 1 \% \rm{,~for~leptons~,} \\
             & = & 80 \% / \sqrt{E} \oplus 5 \% \rm{,~for~hadrons~,}
\end{eqnarray}
where $\oplus$ indicates that the energy dependent and independent
terms are added in quadrature and $E$ is in GeV.

We calculated the $p \bar p $ (for the Tevatron) and $p p$ (for the
LHC) cross sections for the signal with the MRSA$^{\prime}$ structure
functions [34]. We have also examined the effect of using the CTEQ3M [35]
structure functions and found the difference between the two sets of 
structure functions to be small.
We have explicitly calculated backgrounds (1) and (2), and for the
others  used the $Wjj$ background analysis of Ref.[36]. The effect
of the cuts is shown in Table 1.  
Also, in our numerical calculation, we assumed $M_t=175$ GeV,
$\sqrt s=2$ TeV for the upgraded Tevatron and $\sqrt s=14$ TeV for
the LHC. The integrated luminosities for both colliders are assumed
to be 10 fb$^{-1}$.   
 Assuming Poisson statistics,
the number of signal events required for discovery of a signal at
the $95\%$ confidence level is approximately:
\begin{equation}\label{sig}
\frac{S}{\sqrt{S+B}} \ge 3 ~,
\end{equation}
where S (B) is the number of signal (background) events obtained by 
multiplying the signal (background) cross section by the luminosity 
( 10 fb$^{-1}$) and the tagging efficiency for two $b$-jets 
($0.5 \times 0.5$).

With all the above assumptions, we now present the results for
both processes.
\vspace{.5cm}

{\large 3.1~ The B-violating process of $tb$ production}

For the BV process of $tb$ production, 
$q q' \rightarrow {\rm squark} \rightarrow t b$, 
we neglect $ud\rightarrow \tilde s\rightarrow tb$ and
$us\rightarrow \tilde d\rightarrow tb$ since $\lambda''_{112}<10^{-6}$ [10].
For simplicity, we assume $\lambda''_{332}$ and $\lambda''_{331}$ 
do not coexist and 
hence evaluate $cd\rightarrow \tilde s\rightarrow tb$ and
$cs\rightarrow \tilde d\rightarrow tb$ seperately.

Assuming $\Gamma_{\tilde s}=M_{\tilde s}/5$, we obtain 
Fig. 1 which shows the value of $\lambda''_{212} \lambda''_{332}$
versus strange-squark mass for $c d \rightarrow \tilde s \rightarrow t b$ 
to be observable at 95\% confidence level.
The region above each curve is the corresponding observable region.
The solid curve is for the LHC,
the dotted curve is for the upgraded Tevatron and
the dashed line is the perturbative unitarity bound [4,5]. Here we see
that both the LHC and the upgraded Tevatron can efficiently probe
the relevant couplings, and the LHC serves a more powerful probe than
the upgraded Tevatron. 

As was discussed in the above section, 
the width of a down-type squark depends on many free parameters,
which can vary in a large range.
In order to show the sensitivity of the results to 
the width of strange-squark, we present in Fig.2 
the value of $\lambda''_{212} \lambda''_{332}$
versus the ratio $\Gamma_{\tilde s} /M_{\tilde s}$ for 
$c d \rightarrow \tilde s \rightarrow t b$ 
to be observable at 95\% confidence level. Here we assume 
$M_{\tilde s}=300$ GeV.
The region above each curve is the corresponding observable region.
The solid curve is for the LHC and  
the dashed curve is for the upgraded Tevatron. 
We see from this figure that the value of $\lambda''_{212} \lambda''_{332}$
varies mildly as a function of $\Gamma_{\tilde s}/M_{\tilde s}$.
Again it is shown in this figure that 
 the LHC is more powerful than the upgraded Tevatron in probing
the relevant couplings. 

The value of $\lambda''_{212} \lambda''_{331}$
versus down-squark mass for $c s \rightarrow \tilde d \rightarrow t b$ 
to be observable at 95\% confidence level is shown in Fig.3.
The region above each curve is the corresponding observable region.
The solid curve is for the LHC,
the dotted curve is for the upgraded Tevatron and
the dashed line is the perturbative unitarity bound. 
The behaviour of this figure is similar to Fig.1.
But for equal squark mass the value of $\lambda''_{212} \lambda''_{331}$
in Fig.3 is higher than the value of $\lambda''_{212} \lambda''_{332}$
in Fig.1. This shows that the process 
$c s \rightarrow \tilde d \rightarrow t b$  
cannot be probed as efficiently  as  $c d \rightarrow \tilde s \rightarrow t b$
because of the relative suppression of the strange quark structure function
compared to the valence down quark.
\vspace{.5cm}

{\large 3.2~ L-violating process of $t\bar b$ production}

For the LV process of $t\bar b$ production, 
$q \bar q' \rightarrow {\rm slepton} \rightarrow t \bar b$, 
we only consider the dominant process 
$u\bar d\rightarrow {\rm slepton}\rightarrow t\bar b$ and thus
provide the results for 
$\lambda'_{111}\lambda'_{133}+\lambda'_{211}\lambda'_{233}
+\lambda'_{311}\lambda'_{333}$.  
The previous study [30] of this process at the upgraded Tevatron has 
shown that within the  allowed range of the relevant coupling constants,
this process is observable only when the slepton mass lies
in a specific narrow range. Here we will determine if the LHC can do better
than the upgraded Tevatron. 

As discussed in the above section,  the allowed $\lambda^{\prime}$ couplings
which induce a charged slepton to decay into quark pairs are quite small 
and thus the dominant decays of  a charged slepton 
are the chargino and neutralino modes.
So we only consider the chargino and neutralino modes for simplicity.  
Then the width of the charged slepton only depends on the SUSY 
parameters $M_2$, $M_1$, $\mu$ and $\tan\beta$.
In our calculation we use the GUT relation 
$M_1=\frac{5}{3}\frac{g'^2}{g^2} M_2\approx \frac{1}{2}M_2$, 
and fix $M_2=-\mu=250$ GeV and $\tan\beta=2$. We checked 
that in this case the chargino and neutralino masses are 
above the present lower limits from LEP II[37].

Figure 4 shows the value of $\lambda'_{111}\lambda'_{133}
+\lambda'_{211}\lambda'_{233}
+\lambda'_{311}\lambda'_{333}$ versus the slepton mass for 
$u \bar d \rightarrow {\rm slepton}\rightarrow t \bar b$ 
to be observable at 95\% confidence level.
The region above each curve is the corresponding observable region.
The solid curve is for the LHC,
the dotted curve is for the upgraded Tevatron and
the dashed line is the value obtained by considering the 
following bounds for squark mass of
100 GeV [7,11,17]
\begin{eqnarray}\label{eq25}
\vert \lambda^{\prime}_{i11}\vert&<&0.012,~(i=1,2,3),\\
\vert \lambda^{\prime}_{133}\vert&<&0.001,\\
\vert \lambda^{\prime}_{233}\vert&<&0.16,\\
\label{eq28}
\vert \lambda^{\prime}_{333}\vert&<&0.26.
\end{eqnarray}
Figure 4 shows that  below the present upper limit for the couplings the LHC
cannot do much better than the  upgraded Tevatron in further probing
the couplings.  
\vspace{.5cm}

In summary,  we have studied single top quark production via 
$q q' \rightarrow{\rm squark}\rightarrow tb$ and 
$q \bar q'\rightarrow {\rm slepton}\rightarrow t\bar b$
at the Tevatron and the LHC in the MSSM with R-parity violation. 
Our results show that from the measurement of single top production,
the LHC can efficiently  probe the relevant R-parity violating couplings.
 
\vspace{1cm}

\begin{center}{\Large  Acknowledgements}\end{center}

We would like to thank M. Hosch for help in evaluating some of the
backgrounds, and A. Datta and A. P. Heinson for helpful discussions.
X.Z. would like to thank the Fermilab theory group for the hospitality
during the final stage of this work.

This work was supported in part by the U.S. Department of Energy, Division
of High Energy Physics, under Grant Nos. DE-FG02-91-ER4086
DE-FG02-94ER40817, and DE-FG02-92ER40730.
XZ was also supported in part by National Natural Science Foundation of China
and JMY acknowledges the partial support provided by the Henan Distinguished
Young Scholars Fund.
\eject

{\LARGE References}
\vspace{0.3in}
\begin{itemize}
\begin{description}
\item[{\rm[1]}] H1 Collab., C. Adloff et al., DESY 97-024;
                 Zeus Collab., J. Breitweg et al., DESY 97-025. 
\item[{\rm[2]}]  D. Choudhury and S. Raychaudhuri, hep-ph/9702392;
                  G. Altarelli, J. Ellis, G. F. Guidice, S. Lola and M. L. 
                  Mangano, hep-ph/9703276;
                  H. Dreiner and P. Morawitz, hep-ph/9703279;
                  J. Kalinowski, R. R\"uckl, H. Spiesberger and  P. M. Zerwas,
                  hep-ph/9703288;
                  K. S. Babu, C. Kolda, J. M. Russell and F. Wilczek 
                  hep-ph/9703299;
                  T. Kon and T. Kobayashi, hep-ph/9704221;
                  J. E. Kim and P. Ko, hep-ph/9706387;
                  U. Mahanta and A. Ghosal, hep-ph/9706398. 
                  S. Lola, hep-ph/9706519;
                  M. Guchait and D. P. Roy, hep-ph/9707275;
                  T. Kon, T. Matsushita and T. Kobayashi, hep-ph/9707355;
                  M. Carena, D. Choudhury, S. Raychaudhuri and C. E. M.
                  Wagner, hep-ph/9707458.
\item[{\rm[3]}] J. Erler, J. L. Feng and N. Polonsky, Phys. Rev. Lett. 78, 
                 3063 (1997);
                 D. K. Ghosh, S. Raychaudhuri and K. Sridhar, hep-ph/9608352;
                 D. Choudhury and S. Raychaudhuri, hep-ph/9702392. 
\item[{\rm[4]}] B. Brahmachari and P. Roy, Phys. Rev. D50, 39 (1994). 
\item[{\rm[5]}]J. L. Goity and M. Sher, Phys. Lett. B346, 69 (1995). 
\item[{\rm[6]}] F. Zwirner,  Phys. Lett. B132, 103 (1983). 
\item[{\rm[7]}] S. Dimopoulos and L. J. Hall, Phys. Lett. B207, 210 (1987);
                R. M. Godbole, P. Roy and X. Tata, Nucl. Phys. B401, 67 (1993). 
\item[{\rm[8]}] R. N. Mohapatra, Phys. Rev. D34, 3457 (1986);
                M. Hirsch, H. V. Klapdor-Kleingrothaus, S. G. Kovalenko,
                Phys. Rev. Lett. 75, 17 (1995);
	        K. S. Babu and R. N. Mohapatra, Phys. Rev. Lett. 75, 2276 (1995). 
\item[{\rm[9]}] V. Barger, G. F. Giudice and T. Han, Phys. Rev. D40, 2978 (1989). 
\item[{\rm[10]}] J. L. Goity and M. Sher, Phys. Lett. B346, 69 (1995).
\item[{\rm[11]}] K. Agashe and M. Graesser, Phys. Rev. D54, 4445 (1996). 
\item[{\rm[12]}] D. Choudhury and P. Roy, hep-ph/9603363.
\item[{\rm[13]}] G. Bhattacharyya and D. Choudhury, Mod. Phys. Lett. A10, 
                 1699 (1995). 
\item[{\rm[14]}] D. E. Kaplan, hep-ph/9703347. 
\item[{\rm[15]}] J. Jang, J. K. Kim and J. S. Lee, hep-ph/9701283.
\item[{\rm[16]}] J. Jang, J. K. Kim and J. S. Lee, hep-ph/9704213.
\item[{\rm[17]}] G. Bhattacharyya, J. Ellis and K. Sridhar,
		  Mod. Phys. Lett. A10,1583 (1995). 
\item[{\rm[18]}] G. Bhattacharyya, D. Choudhury and K. Sridhar, Phys. Lett. B355,
                  193  (1995). 
\item[{\rm[19]}] J. M. Yang, B.-L. Young and X. Zhang, hep-ph/9705341.
\item [{\rm [20]}]  S. Willenbrock and D. Dicus, Phys. Rev. D34, 155 (1986);
                S. Dawson and S. Willenbrock, Nucl. Phys. B284, 449 (1987);
                C.-P. Yuan, Phys. Rev. D41, 42 (1990);
 	        F. Anselmo, B. van Eijk and G. Bordes, 
                Phys. Rev. D45, 2312 (1992);
                R. K. Ellis and S. Parke, Phys. Rev. D46,3785 (1992);
                D. Carlson and C.-P. Yuan, Phys. Lett. B306,386 (1993);
                G. Bordes and B. van Eijk, Nucl. Phys. B435, 23 (1995);
                A. Heinson, A. Belyaev and E. Boos, hep-ph/9509274. 
\item[{\rm[21]}] S. Cortese and R. Petronzio, Phys. Lett. B306, 386 (1993). 
\item[{\rm[22]}] T. Stelzer and S. Willenbrock, Phys. Lett. B357, 125 (1995). 
\item[{\rm[23]}] A. P. Heinson, hep-ex/9605010. 
\item[{\rm[24]}] M. Smith and S. Willenbrock,  Phys. Rev. D54, 6696 (1996);
                S. Mrenna and C.-P. Yuan, hep-ph/9703224. 
\item[{\rm[25]}] A. Datta and X. Zhang, Phys. Rev. D55,2530 (1997);
                K. Whisnant, J. M. Yang, B.-L. Young and X. Zhang, 
                Phys. Rev. D56, 467 (1997). 
\item[{\rm[26]}] C. S. Li, R. J. Oakes and J. M. Yang, 
                 Phys. Rev. D55, 1672 (1997); Phys.Rev.D55, 5780 (1997);
                 hep-ph/9706412; G.-R. Lu et al., hep-ph/9701406
\item[{\rm[27]}] E. H. Simmons, hep-ph/9612402; C.-X. Yue, Y.-P. Kuang
                 and G.-R. Lu, Phys. Rev. D56, 291 (1997). 
\item[{\rm[28]}] For reviews of the MSSM, see, for example,
                 H. E. Haber and G. L. Kane, Phys. Rep. 117, 75  (1985);
                 J. F. Gunion and H. E. Haber, Nucl. Phys. B272, 1  (1986). 
\item[{\rm[29]}] For reviews of R-parity violation, see, for example,
                G. Bhattacharyya, Nucl. Phys. Proc. Suppl. 52A,  83 (1997).
\item[{\rm[30]}] A. Datta, J. M. Yang, B.-L. Young and X. Zhang, hep-ph/9704257,
                  to appear in Phys. Rev. D.  
\item[{\rm[31]}] C. Carlson, P. Roy and M. Sher, Phys. Lett. B357, 99 (1995);
		 A. Y. Smirnov and F. Vissani, Phys. Lett. B380, 317 (1996). 
\item[{\rm[32]}] H. Baer, A. Bartl, D. Karatas, W. Majerotto and X. Tata, 
                 Int. J. Mod. Phys. A4, 4111 (1989).
\item[{\rm[33]}]  D. Amidei and C. Brock,``Report of the TeV2000 Study Group on
                  Future ElectroWeak Physics at the Tevatron", 1995.
\item[{\rm[34]}] A.D. Martin, R.G. Roberts and W.J. Stirling,
                 Phys. Lett. B354, 155(1995). 
\item[{\rm[35]}]  H. L. Lai, J. Botts, J. Huston, J. G. Morfin, J. F. Owens,
                  J. W. Qiu, W. K. Tung and H. Weerts, Phys. Rev. D51, 4763 
                  (1995).
\item[{\rm[36]}] M. Hosch, private communication.
                 The $Wjj$ background was evaluated using the VECBOS monte
                 carlo program, ( see F. A. Berends, H. Kuijf, B. Tansk
                 and W. T. Giele, Nucl. Phys. B357, 32 (1991) )
                 while the $Wb\bar b$ and $t\bar t$ backgrounds were 
                 evaluated using the ONETOP monte carlo ( see E. Malkawi and
                 C.-P. Yuan, Phys. Rev. D50, 4462 (1994);
                 D. O. Carlson and C.-P. Yuan, Phys. Lett. B306, 386 (1993).)
 \item[{\rm[37]}] B. Mele, hep-ph/9705379.
\end{description}
\end{itemize}
\eject

\begin{table}
\caption{ }
Signal and background cross sections in units of fb after
various cuts at the Tevatron and the LHC. 
The $q q^\prime \rightarrow\tilde q \rightarrow tb$ results have been 
calculated using the unitarity bound for the relevant couplings 
and assuming $M_{\tilde q}=500$ GeV and 
$\Gamma_{\tilde q}=M_{\tilde q}/5$.
 The $q \bar q^\prime \rightarrow \tilde l \rightarrow t \bar b$ 
results have been calculated using the present upper
bound Eqs.(\ref{eq25})-(\ref{eq28}) for the relevant couplings and 
slepton mass of 300 GeV. The slepton width is calculated by
assuming $M_2=-\mu=250$ GeV and $\tan\beta=2$.
The charge conjugate channels are included. 
\vspace{0.1in}

\begin{center}
\begin{tabular}{|c|c|c|c|}
\hline \hline  
Tevatron & basic cuts & basic+m(bW) cuts & basic+m(bW)+bb-tag
\\ \hline \hline
$cd\rightarrow \tilde s\rightarrow  tb$ & 1545 & 1436 & 359  \\ \hline
$cs\rightarrow \tilde d\rightarrow  tb$ & 186  & 174 &  44 \\ \hline
$u\bar d\rightarrow \tilde l\rightarrow  t\bar b$ & 75   &  73 &  18 \\ \hline
$q\bar q^\prime \rightarrow t\bar b$ & 78 & 75 & 19 \\ \hline
$gq\rightarrow q^\prime t\bar b$  & 4 & 3.4 & 0.85 \\ \hline
$qb\rightarrow q^\prime t$ & 236 & 224 & 0.45 \\ \hline
$Wb\bar b$ & 264 & 122 & 30 \\ \hline
$Wjj$ & 62900 & 45000 & 0.7 \\ \hline
$t\bar t$ & 16  & 7 & 1.8 \\ \hline\hline
LHC & basic cuts & basic+m(bW) cuts & basic+m(bW)+bb-tag
\\ \hline \hline
$cd\rightarrow\tilde s\rightarrow  tb$ & 335600 & 304800 &  76200 \\ \hline
$cs\rightarrow\tilde d\rightarrow  tb$ & 125200 & 113600 &  28400 \\ \hline
$u\bar d\rightarrow\tilde l\rightarrow t\bar b$ &482 & 480 & 120   \\ \hline
$q\bar q^\prime \rightarrow t\bar b$ & 573 & 547 & 137 \\ \hline
$gq\rightarrow q^\prime t\bar b$  & 1104 & 810 & 203 \\ \hline
$qb\rightarrow q^\prime t$ & 14150 & 13440 & 27 \\ \hline
$Wb\bar b$ & 756 & 338 & 85 \\ \hline
$Wjj$ & 623600 & 379000 & 6 \\ \hline
$t\bar t$ & 1840  & 644 & 161 \\
\hline \hline
\end{tabular}
\end{center}
\end{table}
\vfill
\eject

\begin{center}{\large Figure Captions }\end{center}
      
Fig. 1 The value of $\lambda''_{212} \lambda''_{332}$
versus strange-squark mass for $c d \rightarrow \tilde s \rightarrow t b$ 
to be observable at 95\% confidence level.
The region above each curve is the corresponding observable region.
The solid curve is for the LHC,
the dotted curve is for the upgraded Tevatron and
the dashed line is the perturbative unitarity bound [4,5]. 

Fig. 2 The value of $\lambda''_{212} \lambda''_{332}$
versus the ratio $\Gamma_{\tilde s} /M_{\tilde s}$ for 
$c d \rightarrow \tilde s \rightarrow t b$ 
to be observable at 95\% confidence level.
The region above each curve is the corresponding observable region.
The solid curve is for the LHC and 
the dashed curve is for the upgraded Tevatron.

Fig. 3 The value of $\lambda''_{212} \lambda''_{331}$
versus down-squark mass for $c s \rightarrow \tilde d \rightarrow t b$ 
to be observable at 95\% confidence level.
The region above each curve is the corresponding observable region.
The solid curve is for the LHC,
the dotted curve is for the upgraded Tevatron and
the dashed line is the perturbative unitarity bound [4,5].

Fig. 4 The value of $\lambda'_{111}\lambda'_{133}+\lambda'_{211}\lambda'_{233}
+\lambda'_{311}\lambda'_{333}$ versus the slepton mass for 
$u \bar d \rightarrow \tilde l \rightarrow t \bar b$ 
to be observable at 95\% confidence level.
The region above each curve is the corresponding observable region.
The solid curve is for the LHC,
the dotted curve is for the upgraded Tevatron and
the dashed line is the present bound, Eqs.(\ref{eq25}-\ref{eq28}). 

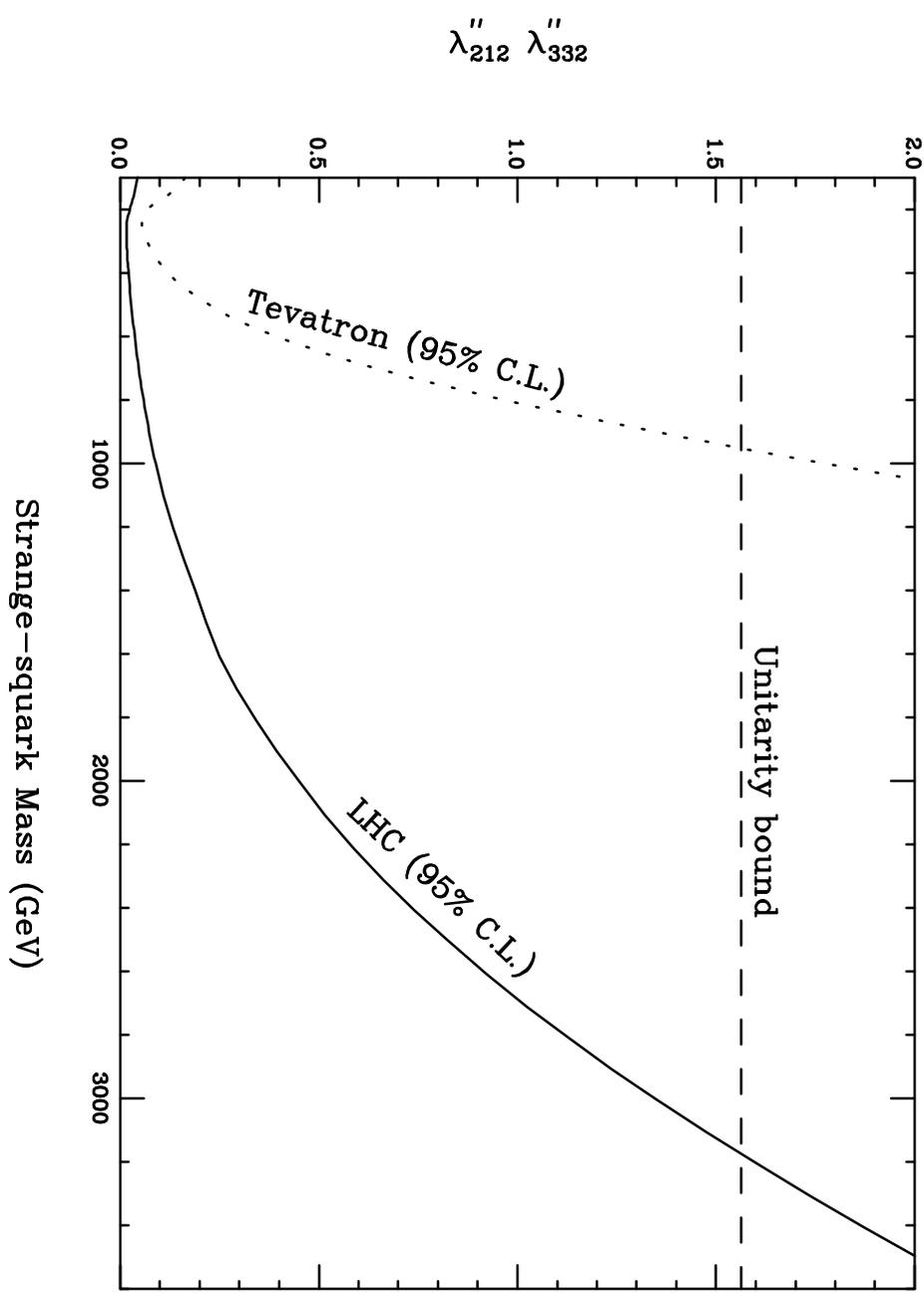
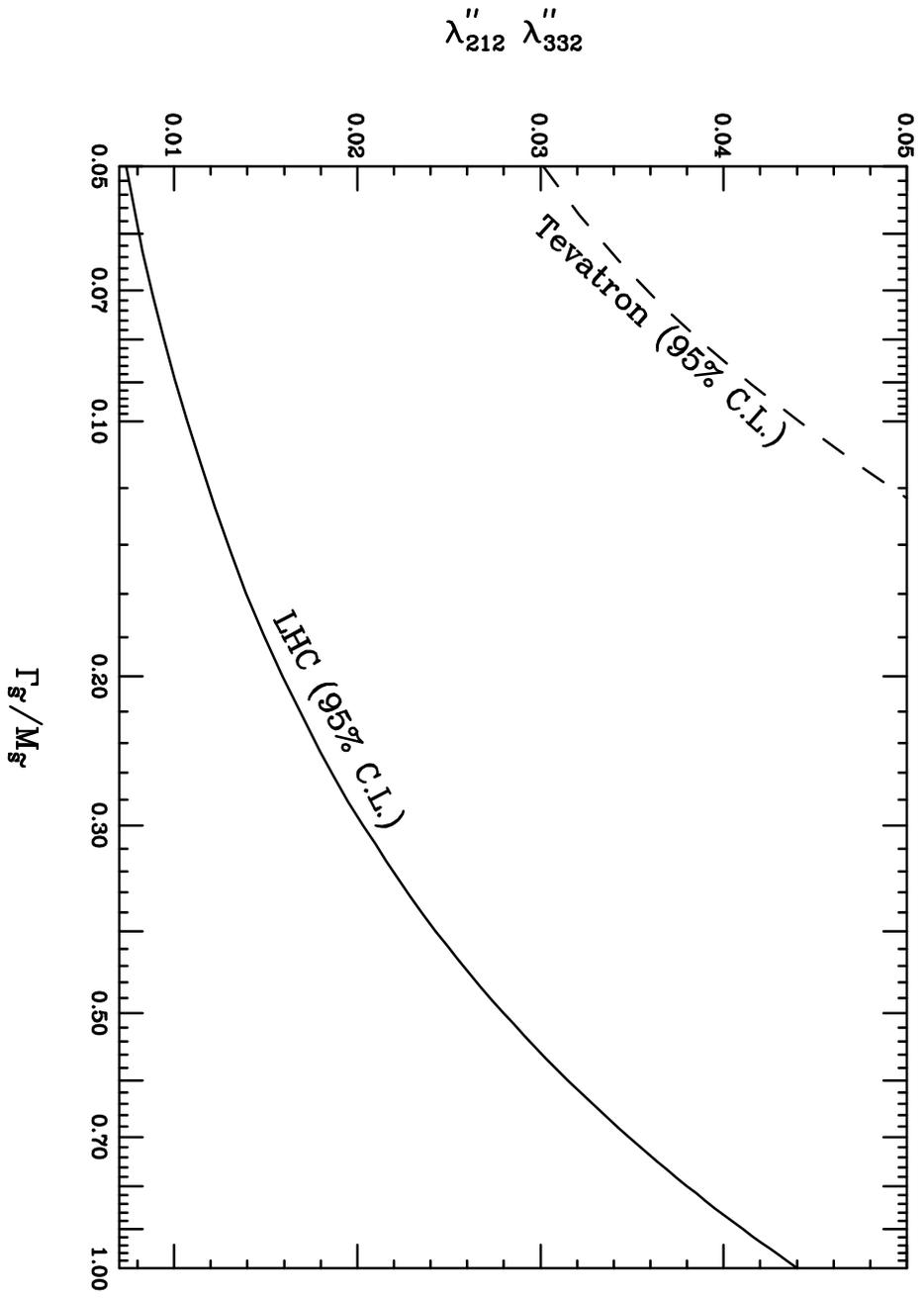
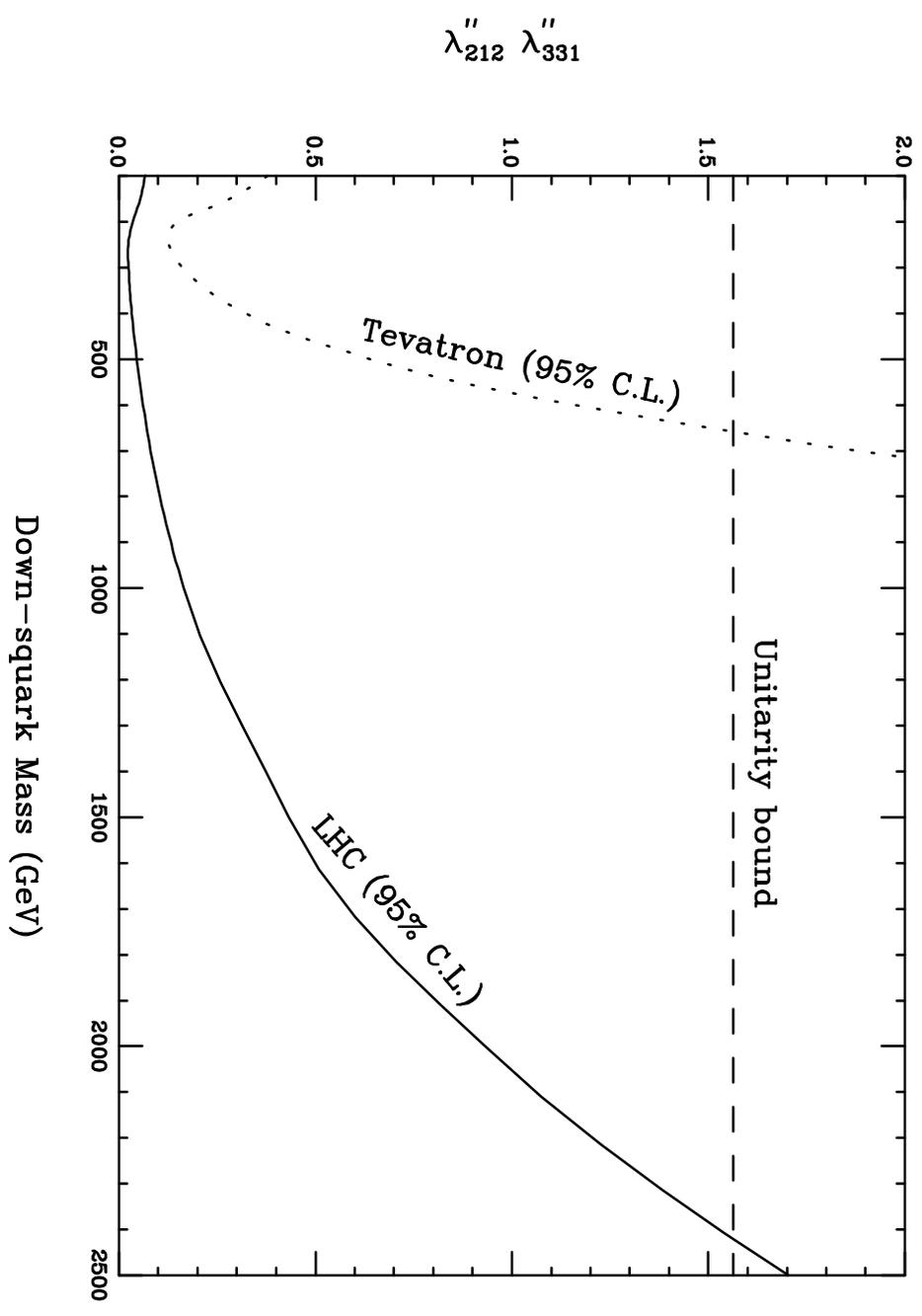
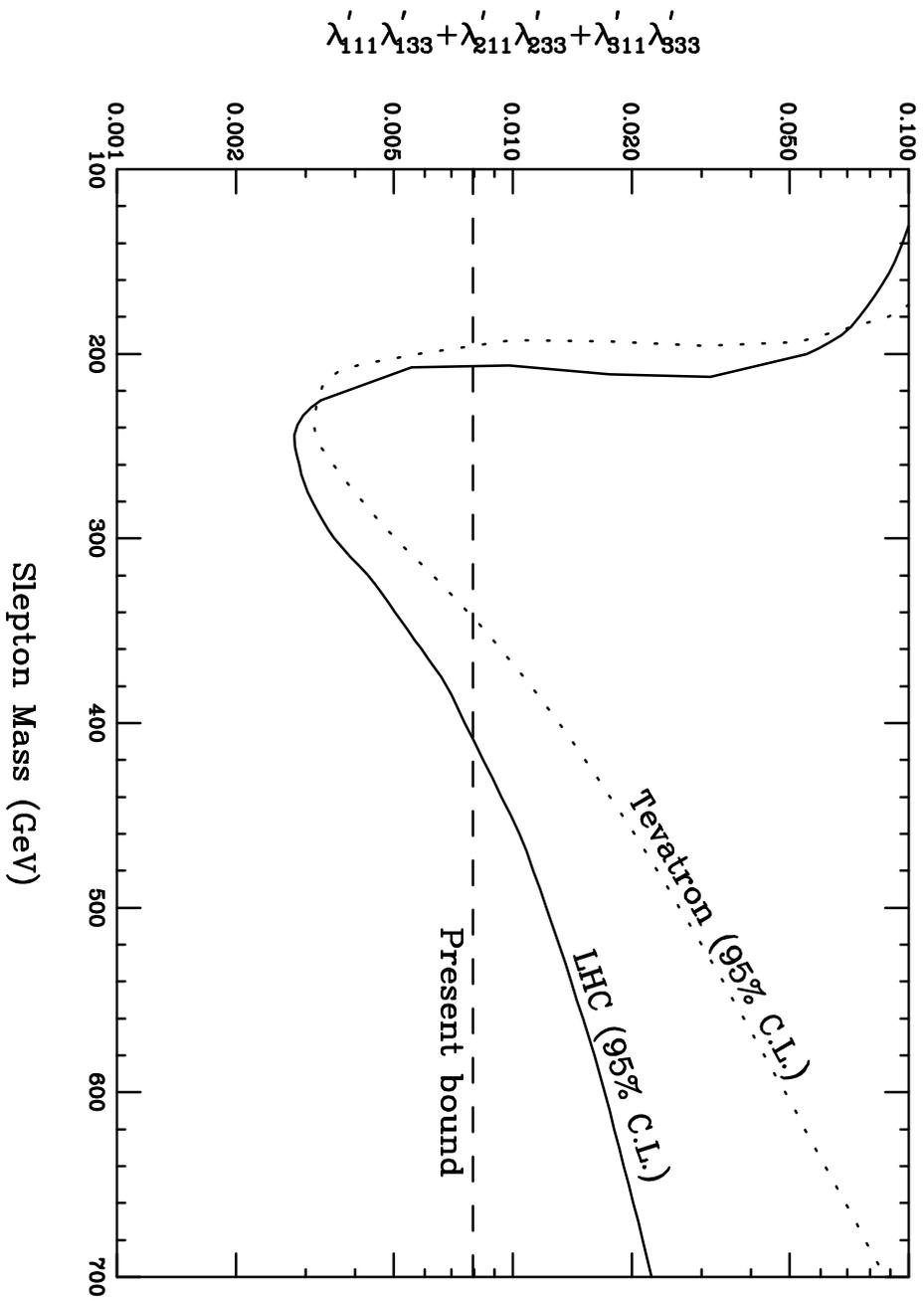
\end{document}